\documentstyle{l-aa}
\begin{document} 
\thesaurus{}
\title
{ The Composite Luminosity Function of A496
\thanks{\rm Based on observations made at the European Southern
Observatory (ESO), La Silla, Chile}}
\author
{Emilio Molinari$^{1}$, Guido Chincarini$^{1,2}$,
Alberto Moretti$^{1}$, Sabrina De Grandi$^{1}$}
\institute{
Osservatorio Astronomico di Brera,
via Bianchi 46, I-23807 Merate (LC), Italy
\and
Universit\`a degli Studi,
via Celoria 16, Milano, Italy \\
e-mail to: molinari@merate.mi.astro.it
}

\date{ Accepted ...; Received ...}
\def\etal{\hbox{et~al.}}
\def\ie{\hbox{i.e.}}
\def\eg{\hbox{e.g.}}
\def\cf{\hbox{cf.}}
\def\kpc{\hbox{~h$^{-1}$~kpc}}
\def\EW{\hbox{W$_\lambda$}}
\def\la{\mathrel{\hbox{\rlap{\hbox{\lower4pt\hbox{$\sim$}}}\hbox{$<$}}}}
\def\ga{\mathrel{\hbox{\rlap{\hbox{\lower4pt\hbox{$\sim$}}}\hbox{$>$}}}}
\def\arcmin{\hbox{$^\prime$}}
\def\arcsec{\hbox{$^{\prime\prime}$}}

\maketitle
\markboth{Molinari et al.: Composite LF in A496}{}

\begin{abstract}
Deep photometric observations in three colours of the cluster A 496 show
that
the 
luminosity function is bimodal with a deep gap at $g \sim 19.0$. That is
there is a 
net separation between E/SO galaxies that are nicely fitted by a gaussian 
distribution curve and the dwarfs that better match a Shechter Function.
This is 
the first cluster observed and reduced out of a sample of 19 clusters 
which we have in our program. However comparison with the data of Virgo and
Coma 
might suggest a correlation between cluster morphology and amplitude of the
two 
distribution: Normal and dwarf population. This would have strong
implication 
for the understanding of cluster formation and evolution so that we are
pursuing 
the estimate of the LF in various colours and to faint magnitudes both for
low 
and high redshift clusters.
\end{abstract}


\section{INTRODUCTION}

A break through in the cluster galaxies luminosity function came with the 
operation of the Du Pont Telescope and the observations carried out by
Sandage 
and collaborators on the Virgo gluster of galaxies (see Binggeli et al. 1988
for 
a review).

\begin{figure}
\centering
\vspace{8cm}
\includegraphics{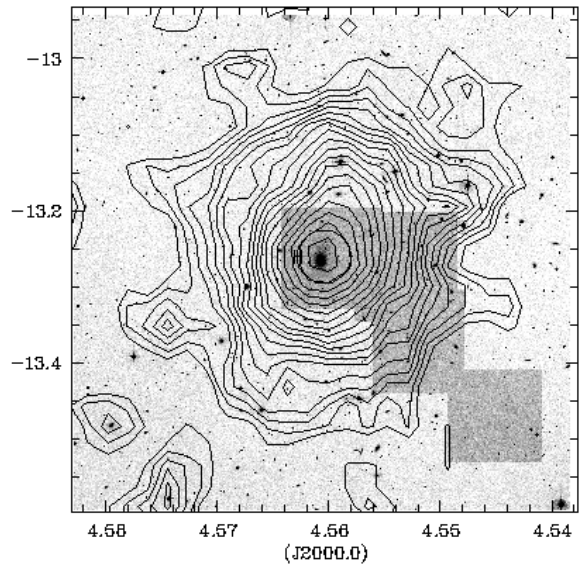}
\caption{
Abell 496 clusters of galaxies image extracted from the POSS digital plates.
The X-ray Rosat all-sky survey contours have been overplotted in arbitrary units.
The circle near the center is the core radius of the X-ray
emission and the shaded area represents the observed fields in this work.
The coordinates refer to the J2000 ephemerides.
}
\label{fig:poss}
\end{figure}

The Virgo cluster is a rather poor cluster and showed that the 
luminosity of normal galaxies follows a gaussian distribution while dwarfs 
galaxies tend to follow a Schechter distribution function. The envelope of
the 
distributions, that is the composite luminosity function of the mix of 
morphological types, is again quite well represented by a Schechter
Function. 
There is no discontinuity between the domain of the normal and dwarfs
galaxies. 
Since the faint end has very strong implication on the cluster mass to 
luminosity ratio and its comparison to the non cluster galaxies give
important 
insights into the cluster formation and evolution, we decided, 1994, to
observe 
the luminosity function of a subsample of clusters detected by the ROSAT 
all-sky survey (De Grandi, 1996). The subsample covers the Bautz-Morgan morphological
types  I, 
I-II and II and will enable us to estimate properties as a function of the 
optical and X-ray cluster morphology. Details of the sample selection, 
properties of the clusters and a description of the data reduction and
analysis 
procedure will be given in Moretti et al (1998). Our study was also
motivated by 
the finding, Marzke et al. 1994, that the non cluster galaxies luminosity 
function raises considerably at faint magnitudes. The comparison cluster
non-
cluster galaxies became mandatory.

After our program started Bernstein et al. (1995) published their very deep 
observations of the central part of the Coma Cluster. These observations are
a 
milestone regarding the faint end of the luminosity function where we wants
to 
disentangle the dwarfs and spheroidal galaxies population from the globular 
clusters. This in order to fully understand the formation and evolution of
that 
population. These authors, however, do not have enough accuracy and
statistics 
on the bright end in order to determine accurately the luminosity function. 
Biviano et al. (1995), on the other hand, using a very clean (low back-for 
ground contamination) spectroscopically selected sample of galaxies in the
Coma 
Cluster show that the luminosity function of that cluster is bimodal. They
notice a lack of galaxies at about $b \sim 17.5$ and that is the separation
between normal and dwarf E/SO. Lopez \& Yee (1996) and Molinari \& Smareglia
(1998) stress that the composite nature of the luminosity function could be a
function of morphology and age given the variation of the galaxy mix as a
function of type. Indeed the fact that the composite (without morphological
type selection) luminosity function in Coma shows a discontinuity while that is
unnoticeable in the Virgo Cluster might call for a strong correlation with
cluster morphology. The claim of an universal form for the cluster luminosity
function (Colless 1989) might come to an end, though it had appealing
connections with a standard fragmentation scenario for structure formation
(see, however, Trentham 1997 and 1998, which find a similar LF both in clusters
and in the field).

We think however that the new findings which we describe here (on
line with the work of Biviano et al. 1995) shed new light on the mechanism of
cluster formation and evolution.

\begin{figure}
\centering
\vspace{9cm}
\includegraphics{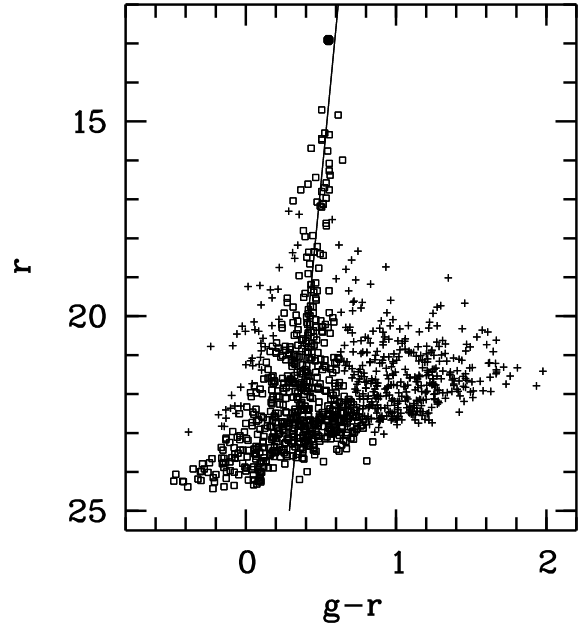}
\caption{
The colour-magnitude plot for the 1058 non star objects
for which we have $g$ and $r$ magnitudes. The
cD galaxy is represented by the larger symbol in the upper part of the plot.
The bright E/S0 galaxies, open squares above $r \sim 18$, fix the solid line
drawn in the middle of the plot and define the c-m sequence for elliptical
galaxies. The population of fiducial early type galaxies which lie on the
c-m sequence is plotted also as open squares (see beginning of section 3).
The distribution of points clearly shows the effect due to the limiting
magnitude in the various colours. Note in addition that below the horizontal
line of $r=20.75$ there is a 10\%-15\% contamination due to galactic stars.
}
\label{fig:cmr}
\end{figure}

\section{THE DATA}

In this first paper we estimate the population content of the cluster A 496 and
measure the luminosity function in three passbands. Abell 496 is a cD dominated
cluster with a very regular, spherical X-ray emission. The hot gas extension
reaches, in the Rosat all-sky survey, 15-20 arcmin from the center. Fig. \ref{fig:poss} shows
the X-ray contours over the digitized POSS image. The gray area has been imaged
in the Dec. 94 run.

The cluster of galaxies A496 is at a redshift z=0.0320. The distance modulus
of the cluster is $m_r-M_r=36.42$ adopting $H_0 = 50 $km/sec/Mpc and a
$k$-correction of 0.01 mag.
To image the cluster we used the DFOSC camera at the Cassegrain Focus of the 
ESO-Danish 1.5 m telescope. In order to sample the cluster at a rather large
distance 
from the cluster center and minimize the observing time, we observed a
mosaic 
area composed of 4 fields along a radial direction and each extended
8.68$\times$8.68 
arcmin$^2$.  The total area observed amounts, accounting for some
overlapping, to 
about 225 arcmin$^2$. Each field has been observed in the colour $g$, $r$
and
$i$ defined 
by the Gunn filters. The centers of the effective transmission, which is 
inclusive of telescope and camera response, are at 4930, 6500 and 8200 \AA.
The 
integration lasted 1 hour per field with a typical seeing of 1.2 arcsec FWHM 
estimated on the stacked images. A pixel size of 0.5" is good enough to
stack 
the shift and stare images as to fully use of the strategy used to gain flat 
field smoothness and ease cosmic rays removal.

Overlapping of different fields allowed us to determine the internal
consistency 
of photometry. We estimated a 1-sigma error at magnitude 22 of 0.20, 0.19
and 
0.22 mag in the $g$, $r$ and $i$ filters respectively. The overall errors
are 
dominated by the poissonian counting noise.

We estimated the effect of crowding in the detection of objects by the
search 
algorithm that we used (package INVENTORY, see West \& Kruszewski, 1981) by 
simulation. Simulated galaxies (extracted by the same CCD images of our
sample) 
of different magnitudes  were added randomly to the images we were measuring
and 
the search process for objects was repeated various times. 
The result is a relation which estimates the probability of detection as a
function of magnitude and position in the field of view: the major source of
incompleteness still remains the background fluctuation, which affects in a
different way the central cD dominated area. Full details on the $p(m,R)$,
where $m$ is the magnitude and $R$ the distance from the cD centre will be
given in Moretti et al (1998).
We measured that we 
had still an efficiency as high as 50\% down to $r \sim 24.5$ all over the
field but 
in the central region. The higher central background light due to the cD
galaxy 
prevented us to detect objects fainter than $r \sim 23$. 
The area involved is nevertheless limited and the cD halo action in hiding
faint objects starts its efficiency at $r > 23$.
Part of the problem of the 
detection of faint objects in the presence of a bright background due to the 
halo of bright galaxies was overcome by modelling the shape of the giant 
galaxies and removing them from the frames to be measured. Details on the 
complete data reduction procedures will be given by Moretti et al. (1998) in 
conjunction with the whole photometric catalogue.

Up to $r \sim 20.5$ we were able to distinguish easily stars from galaxies
using the 
isophotal radius. Fainter objects were however seeing limited so that a safe 
estimate based on the PSF was not possible. 
We thus decided to keep in a conservative way all the faint objects in our
working catalogue. An analysis of the galactic stellar contamination (Robin
et al. 1995) yielded a number of $\sim 300$ stars with magnitude $20.75 < r
<
24.4$, or about 15\% of our objects. Restricting the analysis where the
sequence of early type populates the c-m plane (see Fig. \ref{fig:cmr}) the
level of contamination is even lower (less than 10\%).
All this taken into account, from the complete photometric catalogue of 2355 
objects
a set of 279 identified stars with $r<20.75$ are removed. The remaining
2076 objects, with at least one filter measured, 
suffer from contamination between $r$ = 20.75 and
the limiting magnitude of $r \sim 24.5$ from the presence of a 10\%-15\% of
galactic stars. As the stars are rather few, the results of the analysis
about the luminosity functions should not be altered by such low
contamination. In the case of correction by subtraction the star
contribution is statistically cancelled, and the study of the E/S0
population
(see Section 4.1) is  affected by a even lower star fraction.

We were able to measure $g$ magnitudes for 1080 galaxies, $r$ for 2054 and 
$i$ for 1499: 1058 galaxies have both $g$ and $r$ magnitudes and 955 $g$,
$r$ and $i$.
The limiting magnitudes (50\% efficiency in detection) are 24.14, 24.46,
23.75 for $g$, $r$ and $i$ respectively.
All the magnitudes used in this work are isophotal magnitudes, computed
integrating the flux within an isophote of $2\sigma$ the sky noise in each
filter. For ease of comparison with other magnitude definition we have
undergone the whole photometric measurement for a set of objects with known
{\it true\/} flux, obtained scaling down to faint brightness a bright, non
saturated object (much like as used in the incompleteness analysis). Our
isophotal magnitude does not differ significantly up to $r \sim 23.5$ for
stellar like objects from the known magnitude. For an elliptical galaxy
surface brightness distribution the error, in the sense of our magnitude
look fainter, is as much as 0.47 mag when the {\it true\/} one is 22.
Somewhat higher differences are expected for disk like and irregular
galaxies. In our subsequent analysis we have not performed any correction
and used Inventory isophotal magnitudes.

\section{THE CLUSTER POPULATION}

As expected for a cD cluster, early type galaxies dominate the central
region 
and these are easily identified by visual inspection down to $r \sim 18$. By
using 
this set of bona fide early type galaxies we can compute the best fit of the 
linear sequence they define in the $g$-$r$ $/$ $r$ plane and extrapolate it
to fainter 
magnitudes. The linear sequence of early type galaxies is, indeed, well
defined 
down to $r \sim 21.5$ with an intrinsic spread about the mean of ($g$-$r$)
$\sim 0.2$ mag at $r \sim 21$. 
The presence of a distinct population which obeys a colour-magnitude relation 
allowed diffrent author to single out the early type population belonging to 
a cluster (among the others Metcalfe et al. 1994, Secker 1996 for rich clusters,
Molinari and Smareglia 1998 using a neural network to exploit multi-colours
information and De Propris and Pritchet 1998 who extend the c-m down to the 
faint magnitude of $M_I \sim -14$).
We thus decide to partition the whole set into three classes: the fiducial
early type, lying in the region of the $r$ vs. $g$-$r$ plane closer to the
c-m relation less than 1 $\sigma$ of the poissonian error of our photometry
(i.e. $|[(g-r)-cm_{E/S0}]| \le 0.28\times 10^{0.2(r-22)}$ ); the red
objects, with $g$-$r$ colour greater than $\sim 0.8$; the objects bluer than
$g$-$r$ $\sim 0$. 
In Fig. \ref{fig:cmr} 
we mark with different symbols the early type galaxies that belong to the 
linear sequence we mentioned and the bluer and redder objects which define 
different populations.

\begin{figure}
\centering
\vspace{15.8cm}
\includegraphics{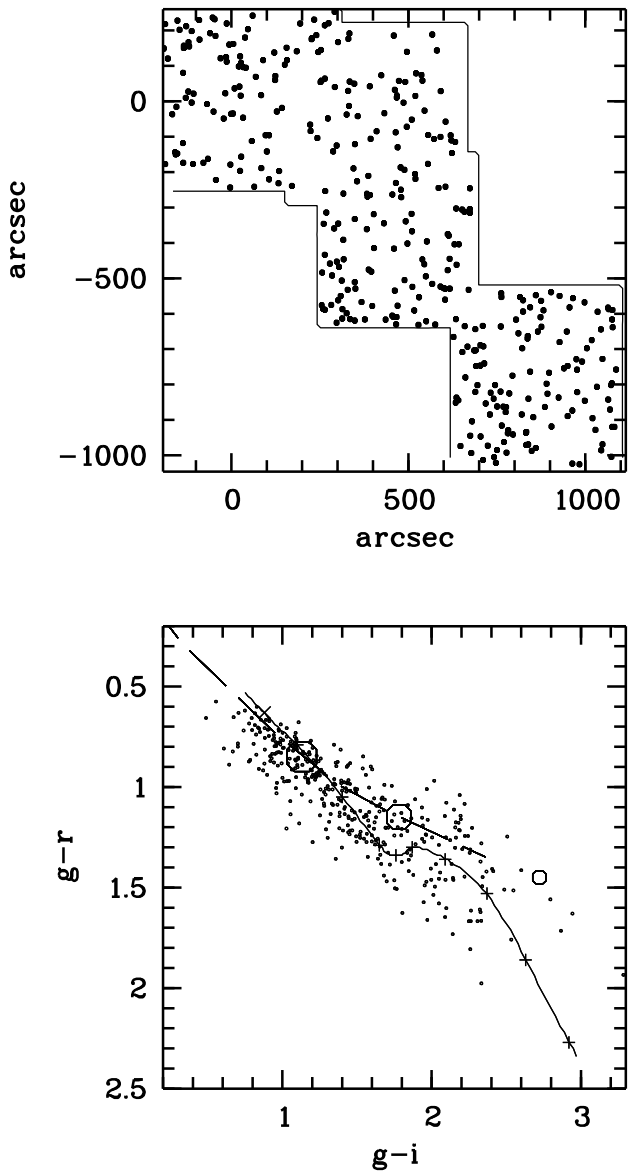}
\caption{
Distribution of the redder objects as described in the text. The upper plot
shows the distribution 
on the sky: this is quite uniform without any evidence of clustering. In the
bottom panel 
the same objects have been plotted in the colour-colour plane, $g$-$r$
versus
$g$-$i$. All the 
points are spread, without any particular clustering, over the line
indicating the position 
of elliptical galaxies as a function of redshift: the crosses are separated
by a difference 
of 0.1 in redshift starting from $z=0$ (upper left) to $z=1$ (bottom right).
The scarcity of 
points going toward the red is expected in a sample limited in magnitude. In
the upper plot 
the origin has been taken at the position of the cD galaxy while the
evolutionary sequence 
in the bottom diagram has been computed using the population synthesis
models by Buzzoni (1995). The dashed line shows the locus of the main
sequence for galactic stars. The circles on the line have an area
proportional to the expected fraction of stars as computed with the Robin et
al. (1995) model of the galaxy, on which basis a total of $\sim 60$ stars
should be present between $r=20.5$ and the completeness limit.
}
\label{fig:red}
\end{figure}

\subsection{The red and the blue objects}

The two extreme populations, i.e. the reddest and bluer objects which lie
apart from the early type c-m relation, deserve at least a brief attention. 

The red objects can be viewed in their spatial distribution in Fig.
\ref{fig:red} (upper panel), showing no clear sign of clumpiness. The random
distribution is evidenced also in the $g$-$r$ vs. $g$-$i$ phase space: the
presence of a fore- or back- ground cluster would have shown up as a
overdensity of points at a particular colour location. Using a population
synthesis model (Buzzoni 1995, the solid line in Fig. \ref{fig:red}, lower
panel) we can follow the expected colour shift as a function of redshift
predicting the effect of a purely passive evolutionary $k$-correction. The
points in the plane clearly stretch from bluer (zero redshift) to redder
colours, spreading out to a redshift $z \sim 0.8$. We are therefore lead to
the classification of these objects as background (mainly E/S0) galaxies.
Some of the points, anyway, are probable galactic stars, especially in the
upper, bluer $g$-$i$, strip (along the dashed line in Fig. \ref{fig:red}
which shows the locus of main sequence stars).

At the left of the c-m relation (Fig. \ref{fig:cmr}) we observe a handful of
blue objects which we believe disk galaxies belonging to the cluster. 
These are in limited number in agreement with the cluster morphology,
Bautz-Morgan I, which calls for a population dominated by early type
galaxies. Always in agreement with the cluster morphology, these galaxies
are located in the cluster outskirts and their density peaks at about 500
arcsecs from the cluster center or at about 1 core radius.

\begin{figure}
\centering
\vspace{7.2cm}
\includegraphics{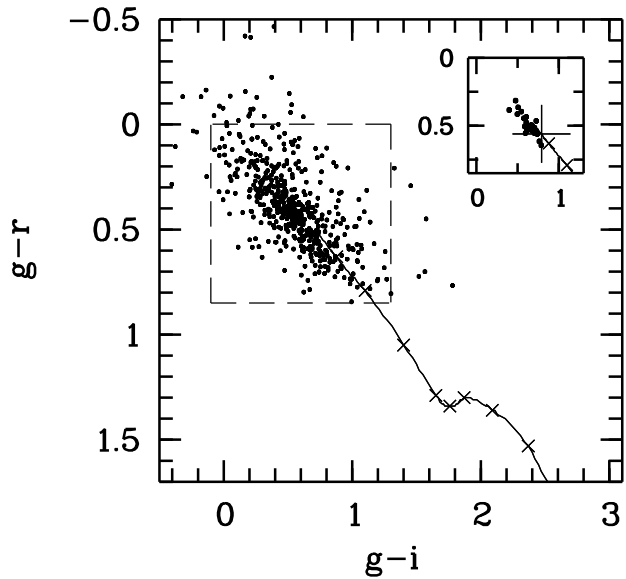}
\caption{
The $g$-$i$, $g$-$r$ plot for the fiducial early type galaxies as identified
in
the c-r sequence. The solid line is the same photometric model (Buzzoni
1995)
as in Fig \ref{fig:red} and the ticks mark every 0.1 in $z$. The inset panel
plots the dashed region selecting only the bright ($r<18$) galaxies showing
a
narrower distribution. With negligible poissonian errors the colour
dependence
on the magnitude is the only source of (intrinsic) scatter. On the place of
modelled colours for a $z=0.03$ galaxy, the big cross closely matches the
brightest galaxies in our sample.
}
\label{fig:ccc}
\end{figure}

\subsection{The early type galaxies}

Their distribution in the colour-colour plot, Fig. \ref{fig:ccc},
match what we would expect for an ensemble of early type galaxies at the
cluster 
redshift. 
The morphological type and the colour-magnitude sequence allowed us to
define
a 
sub-sample of E/S0 galaxies that quite likely are cluster members, indeed
the 
main component of the cluster. 
The cluster of points in the plot is rather scattered. This is due to 
two effects: {\it 1)} an intrinsic scatter in the colour of elliptical
galaxies mainly 
due to the correlation between absolute magnitude and colour and {\it 2)}
measuring 
errors in the colour of the faintest galaxies. For comparison and clarity we 
inserted in Fig. \ref{fig:ccc} the colour-colour plot for the cD and for the
galaxies with 
$r < 18$ magnitude. The large cross marks the expected position a very
bright 
elliptical galaxy would have in this diagram at the cluster redshift. As 
expected due to the cleaner sample and to the more accurate magnitudes of
the 
bright galaxy colours, the plot presents a much smaller scatter.

Granted that the early type galaxies better define the cluster and seem to
be, 
at least in the central part, the cluster itself, we look at their spatial 
distribution. This is not fit by a single King profile and in fact we can
not 
define a regularly decreasing density function by using only these galaxies.
At 
about 1 to 1.5 core radii we detect a rather well defined substructure. To
check for luminosity segregation we
choose to restrict the analysis for the brightmost E/S0 ($r < 20$) finding a
quite regular cluster density distribution with a core radius $R_c = 242$
h$^{-1}$ Kpc (h = $H_0/100$ km/sec/Mpc). 
Also morphological segregation can be quantified by looking at
the density profile of all (not only E/S0) galaxies brighter than $R=20$.
The $R_c$ become larger, as expected, to an extent of $R_c = 330$ h$^{-1}$
Kpc.

\section {THE LUMINOSITY FUNCTION}

\begin{figure}
\centering
\vspace{6cm}
\includegraphics{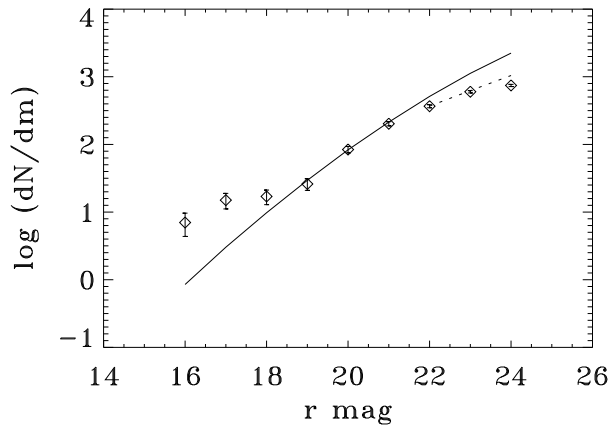}
\caption{
Comparison between Tyson (1988) number counts (solid line). A496 counts are
plotted with poisson error bars. It is easy to recognise the overdensity
due to the cluster presence at bright magnitudes and the incompleteness of
our observations that starts after $m \sim 22$.
The dashed line represents the correction to our counts for the
incompleteness due to crowding and background level differences due to cD
presence. The correction is based on a bootstrap technique as described in
section 2 and only affects significantly magnitude bins $r > 23$.
} 
\label{fig:tyson}
\end{figure}

The best way to subtract the background is to subtract a control sample
obtained 
with the same instrumentation used for the object sample. That is why we 
selected to observe the cluster also at large distances from the cluster
center. 
The furthermost frame of our sample is located at about r = 2.5 cluster
radii using $R_c = 242$ h$^{-1}$ Kpc. Assuming a King profile distribution,
this is quite reasonable for the 
scope, we expect that in this frame the density of cluster galaxies is less
than 
about 1/12 the density we have in the central region of the cluster.  The 
outskirts region is, for our purpose, quite suitable for background number 
counts correction. In fact our counts in the outskirts of the cluster are 
in excellent agreement with Tyson (1988 and references therein). 
In any case our method of subtraction of two nearby (in fact neighbouring)
areas will eliminate some contribution of pure cluster population but, at
the same time, overcomes at best the problem of cosmic variance in
background numbers counts which acts as major error source at the faint end
of LF (Trentham 1997). 
In our analysis we cannot and will not take into account such variance as
error in background subtraction, considered as affected by counting
statistics only.
Figure \ref{fig:tyson} is 
quite illustrative of the work we are doing. Here we plot our rough counts,
that 
is before detailed processing as a function of the distance from the cluster 
center, as a function of magnitude superimposed to the counts by Tyson. 
The incompleteness correction technique described in section 2 can give a
very limited help in the number counts measurements. It only affects
seriously the very faint magnitudes ($r> 23$) even if the contribution of
the hiding cD halo is taken into account (see the dashed line in the faint
magnitude bins in Fig. \ref{fig:tyson}).
We clearly evidence the main features of the sample: {\it 1)} the bump at
the bright 
magnitudes as due to the presence of the cluster core, {\it 2)} over the
whole field 
the cluster is easily washed out at $r \sim 20$ when compared to absolute
counts and 
{\it 3)} from about $r \sim 22$  to $r \sim 24$ magnitude we evidence 
the incompleteness of our sample. 
Obviously this does not affect our work because we took that into account in
two 
different ways: {\it a)} we estimated the incompleteness by simulation as
mentioned in 
Section 2 and detailed in the paper by Moretti et al. (1998) and {\it b)} we
chose to use 
only our frames for subtractions so that the limiting magnitude of the two
operand is exactly the same and the background population are as close as
possible.

\begin{figure}
\centering
\vspace{12cm}
\includegraphics{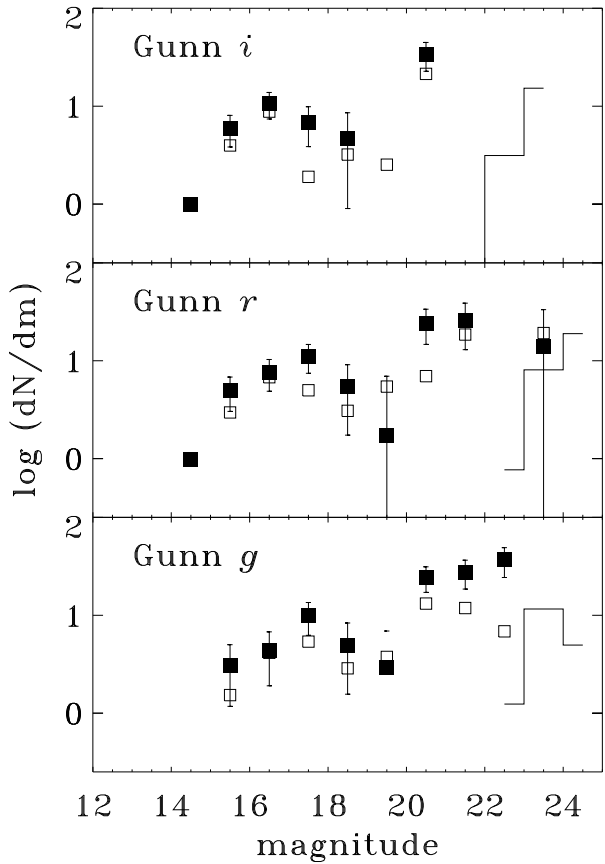}
\caption{
Luminosity function in the filters $g$, $r$ and $i$ for the central
region of A496. The error bars have been computed using poisson statistics.
Filled squares to galaxies within 600 arcsec from the centre
(without error bars to avoid confusion)
while open squares refer to galaxies within 300 arcsec. All the objects
which were not classified as bright stars ($r<20.5$ as described in section
2) are used for the analysis: the distribution within 600 (or 300) arcsec
from cluster centre are corrected subtracting the distribution of the outer
frame properly normalized. The histogram at faint magnitudes in each panel
shows the statistically expected losses of objects due to the presence of
the cD halo.
}
\label{fig:lfin}
\end{figure}

The complete cluster luminosity function, that is using all galaxies 
independently of their colour or morphological type, is simply determined as
the 
difference between the total distribution in magnitudes and that related to
the 
background. Obviously the maximum contrast between cluster and background is
obtained by 
using the central field for the cluster and the outermost field for the 
background. The luminosity function so derived is illustrated in Figure
\ref{fig:lfin} in the 
three different colours in which we observe.

\subsection {The Composite Luminosity Function of A496}

It is clear that in all colours we observe, Figure \ref{fig:lfin}, a minimum
in the 
distribution about the magnitude $m \sim 19$ which corresponds to a net
separation 
between the distribution of giant normal galaxies and dwarfs. This is very 
similar to the result obtained by Biviano et al. (1995), albeit much more
robust because of the more net separation, the three 
colours and the higher statistics. As we are dealing with bright objects no
sensible effect due to the choice of our isophotal magnitude is expected.

We must point out, however, that Biviano et al. were able to evidence 
the effect because of the very clean selection of cluster members due to the 
fact that for the Coma galaxies a large number of redshifts is available.
That 
is important because it means that the detected effect is not so evident in
Coma using counts alone and that is demonstrated by the work by Bernstein et
al. (1995). 

To some extent we can say that such a result was implicit
in the Virgo cluster works by Binggeli et al. (1988), which could identify
morphological types of galaxies, (1988) and Capaccioli et al. (1992), which
found two distinct population among the early type and spiral bulges in the
fundamental plane. 

\begin{figure}
\centering
\vspace{8cm}
\includegraphics{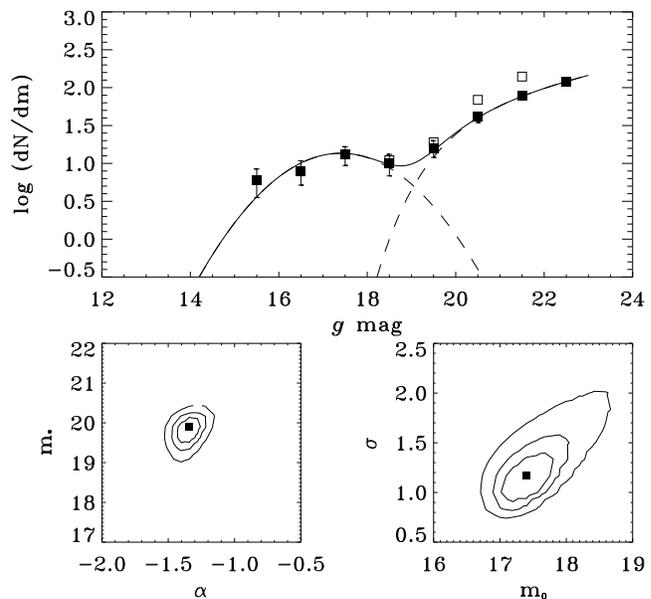}
\caption{
Luminosity function in the filter $g$ for the 
early type galaxies, as selected in the colour-magnitude plane of Fig.
\ref{fig:cmr}. A magnitude cut-off was set to $r=23.0$ to avoid the
incompleteness in the $g$ passband. This selection made the subsample used
for early type galaxies LF determination 90\% (or higher) complete in all
three filters.
The points (and poissonian error bars) are
built from a 1 mag binning and are shown for comparison with the function
derived from maximum likelihood methods. The dashed lines are the gaussian
(leftmost) and Schechter (rightmost) separated distribution which contribute
to
the sum (solid line). For completeness we added, shown as open squares, the
LF with the corrected magnitudes (see section 2): the steepness of the faint
end increases as expected. The lower panels shows $1\sigma$, $2\sigma$ and
$3\sigma$
confidence intervals for the composite function parameters, computed with
the uncorrected isophotal magnitudes. 
}
\label{fig:lfeg}
\end{figure}

\begin{figure}
\centering
\vspace{8cm}
\includegraphics{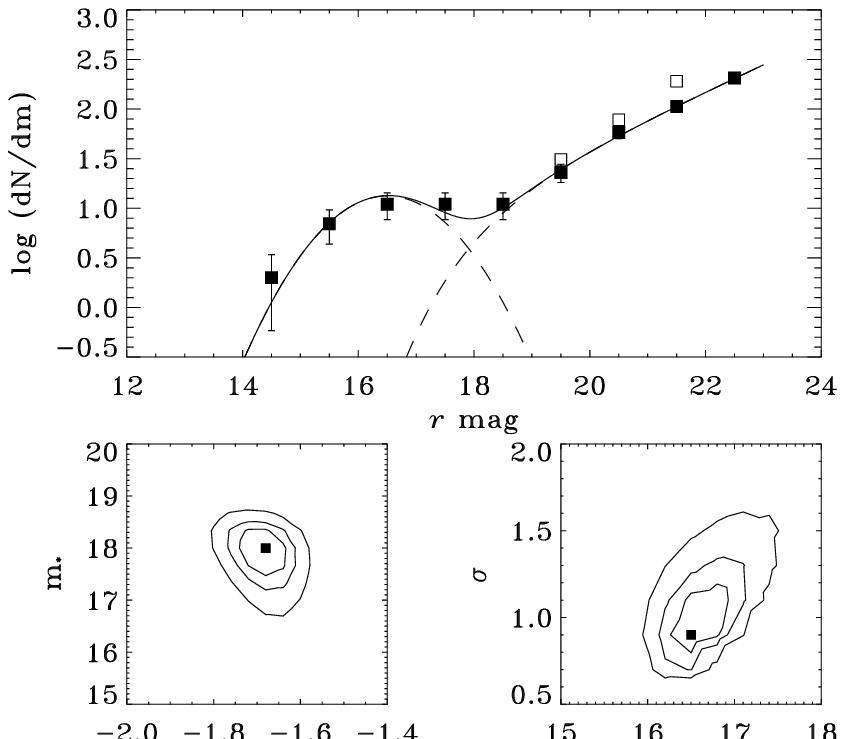}
\caption{
Luminosity function in the filter $r$. Same as Fig. \ref{fig:lfeg}.
}
\label{fig:lfer}
\end{figure}

\begin{figure}
\centering
\vspace{8cm}
\includegraphics{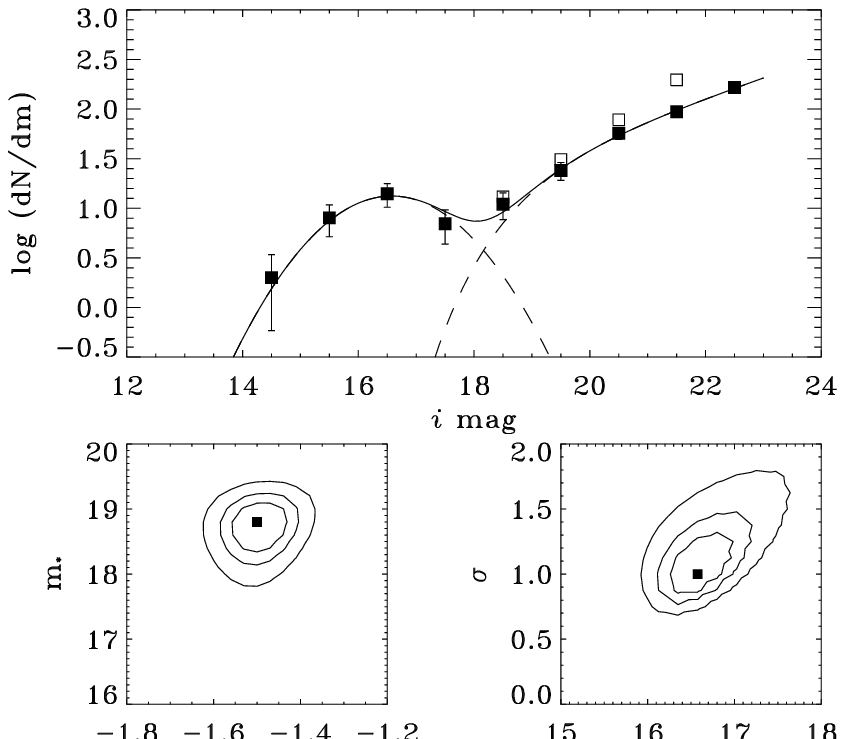}
\caption{
Luminosity function in the filter $i$. Same as Fig. \ref{fig:lfeg}.
}
\label{fig:lfei}
\end{figure}

We then take into consideration the subset defined by the fiducial
early-type galaxies that follows the colour-magnitude relation, as shown in
Fig \ref{fig:cmr}. This sample is quite reasonably composed by member
galaxies and constitutes the major component of the cluster population, as
discussed in section 3.2. Binned counts are shown in Figs. \ref{fig:lfeg},
\ref{fig:lfer}, \ref{fig:lfei} with poisson errors. 
The way to fit the observed counts is self-evident looking at the
distribution. 
We have chosen to fit it by using a gaussian plus a Schechter function. 
Other functions (e.g. Erlang) might obtain a better fit, but it is not clear
however what that would physically mean.
Given the small difference we prefer, at this time, not to look into more
details. We 
used a maximum likelihood method exploring a 5-dimension grid of parameters 
defined by the function:

$$ {dN \over dm} = \phi \left\{ S(\alpha, m_*) + \rho G(m_0,\sigma)\right\}
$$
where
$$ S(\alpha, m_*) = 10^{-0.4(m-m_*)(\alpha+1)}e^{-10^{-0.4(m-m_*)}} $$
and
$$ G(m_0,\sigma) = e^{-(m-m_0)^2/2\sigma^2} $$.

The normalization factor $\phi$ is computed integrating the number of
objects that 
have $r \le 23$. Here we stop the computation of LF due to the
incompleteness of the 
$g$-$r$ colour distribution at these magnitudes. The parameter $\rho$
modulates the ratio 
between the gaussian distribution and the Schechter function. In Table
\ref{tab:lf} we 
report the results and the 1 sigma confidence interval for the 5 parameters 
$\alpha$, $m_*$, $m_0$, $\sigma$ and $\rho$.

In every filter the case of a null contribution from the gaussian function
is 
excluded at a very high confidence level ($>3\sigma$). Its value, however,
differs in 
different colours and this is mainly due to the limited statistics and
fitting 
procedure. That will be improved with other cluster estimates as soon as the 
data analysis will be completed.

This is an important result and definitely indicates a bimodal distribution 
for the cluster luminosity function. Indeed now that we have these results
we 
can look at the sequence which goes from the irregular to regular clusters,
from Bautz-Morgan type III to Bautz-Morgan Type I. We can rely only on three
clusters: Virgo, Coma and A496. The indication is clear however. In the very
irregular Virgo cluster the E/S0 contribution to the luminosity function  is
not so dominant and it is hardly noticeable in the composite luminosity
function. The gaussian peak magnitude is at $b_T$ = 13.0, equivalent to $M_i
\sim -20.5$ (Binggeli et al, 1988, also report of a scarce but similarly
bright component of E/S0 in the field).

The bump is noticed at the same $M_i \sim -20.5$ in the clean sample
selected by Biviano in the Coma cluster. This is not so evident, however, in
the counts used by Bernstein et al. (1995) even if a plateau could be
guessed. On the contrary it is clear and sound in A496, peaking at $M_i \sim
-20$. 
Accordingly the evidence of the lack of galaxies on the right of the
gaussian peak becomes more evident.

On the basis of these preliminary indications we thus expect that the
contribution of the bright gaussian increases with the cluster morphological
type, from irregular to relaxed shapes. 
Again this is a major result having very strong implication in the formation
and evolution of clusters and which we will further test with the analysis
which is still in progress.

The faint end of our A496 luminosity function is slowly rising with a slope
parameter $\alpha \sim 1.65$ for the $r$ band. The effect on the steepness
of a different magnitude definition is visible in Figs.
\ref{fig:lfeg},\ref{fig:lfer} and \ref{fig:lfei} marked with open squares.
The $\alpha $ parameter increases up to $\alpha \ge 2.0$ for the same $r$
band. Similar conclusions can be obtained for all the filters.

\section{Summary and Conclusions}

We have discussed the catalogue listed in Moretti et al (1998). The
catalogue is
the result of measuring about 2000 objects in 4 frames and in three ($g$,
$r$ and $i$) colours. The major result of this work is that we established
that in the cD cluster A496 the luminosity function is bimodal with the
early type population clearly fitted by a gaussian distribution while the
dwarf galaxies obey a Schechter function with a rather high slope on the
faint end.
With an absolute magnitude of $M_r \sim -18.5$ we find the gap 1 mag fainter
than previously observed in Coma (Biviano et al. 1995)
The indication is that the amplitude of the gaussian respect to the
Schechter function depends on the cluster morphology, going from a barely
noticeable gap in irregular clusters (e.g. Virgo) to a clear evidence in
evolved clusters (Bautz-Morgan type I). 

The somewhat high slope in the dwarf domain, with $\alpha \sim 1.65$ does
not differ too much from other authors findings in similar clusters
(Trentham 1997, 1998). While it is not possible to compute the value of
$\alpha$ at $M_r =-20$ because of the gaussian shape of the LF in the bright
end, the comparison between our result and their work give a higher slope
for A496. Taking into consideration the discussed correction for the
magnitudes we even reach the very steep value of $\alpha \sim -2.0$, close
to the value reported by De Propris et al (1995) for the dwarves in A665. 
This thesis will be fully developed in the work that is continuing with the
complete photometry for the 19 selected clusters.

\begin{table}
\centering
\caption{A496 LF parametrization for the $g$, $r$ and $i$ filters.}
\begin{tabular}{c|c|c|c|c|c}
& $\alpha$             & $m_*$                & $m_0$                &
	  $\sigma$            & $\rho$              \\
\hline
$g$& -1.34 $^{+.06}_{-.04}$ & 19.8 $^{+.2}_{-.2}$ & 17.4 $^{+.2}_{-.2}$&
      1.10 $^{+.17}_{-.13}$ & 0.22 $^{+.04}_{-.04}$ \\
$r$ & -1.69 $^{+.04}_{-.04}$ & 18.0 $^{+.3}_{-.3}$ & 16.5 $^{+.2}_{-.2}$&
       3.5 $^{+.00}_{-.20}$ & 1.07 $^{+.18}_{-.20}$ \\
$i$ & -1.49 $^{+.04}_{-.04}$ & 16.0 $^{+3.0}_{-.0}$ & 16.6 $^{+.2}_{-.2}$&
       1.04 $^{+.16}_{-.12}$ & 0.44 $^{+.09}_{-.08}$ \\
\hline
\multicolumn{6}{l}{Errors quote $1\sigma$ confidence intervals}\\
\end{tabular}
\label{tab:lf}
\end{table}

\end{document}